\documentclass[conference]{IEEEtran}
\IEEEoverridecommandlockouts
\usepackage{cite}
\usepackage{multicol}
\usepackage{lipsum,graphicx,subcaption}
\usepackage{float}
\usepackage{epstopdf}
\usepackage{ifthen}
\usepackage[latin1]{inputenc}

\usepackage{amssymb}
\usepackage[cmex10]{amsmath}
\usepackage{dcolumn}
\usepackage{fixltx2e}
\usepackage[nolist]{acronym}

\DeclareGraphicsExtensions{.pdf,.jpeg,.png,.eps}
\usepackage{booktabs}

\usepackage{comment}
\usepackage{color}

\usepackage{setspace}
\usepackage{mathrsfs}          % provides math script font \mathscr{}
\usepackage{amsmath, amssymb, amsthm}
\usepackage{amsfonts}
\usepackage{mathrsfs}
\usepackage{multirow}        % for centering over rows in tables
\usepackage{framed}
\usepackage{psfrag}
\usepackage{units}            % for nicely typset units (m, cm, dB)
\usepackage{algpseudocode} % for algorithm
\usepackage{textcomp}
\usepackage{xcolor}

\newcolumntype{d}[1]{D{.}{\cdot}{#1} }

\usepackage{bm}

\newcommand{\LeftP} {\left\lbrace }
\newcommand{\RightP}{\right\rbrace }
\newcommand{\LeftPD} {\left( }
\newcommand{\RightPD}{\right)}

\newcommand{\compl}{\mathbb{C}}         % complex number field, e.g., x \in \compl
\newcommand{\ma}  [1]{ \bm{#1} } % matrix (upper case) and vector (lower case)
\newcommand{\mav}  [1]{ \bm{#1} } % matrix (upper case) and vector (lower case)
\newcommand{\IndexM}[3]{\left[ #1\right]_{\LeftPD #2,#3 \RightPD } } % indexing Matrix
\newcommand{\IndexV}[2]{\left[ #1\right]_{\LeftPD #2 \RightPD } } % indexing vector

\newcommand{\dft} [1] {\tilde{#1} } % Dft of vector or matrix

\newcommand{\Ex}[1]{\mathrm{E}\left[ #1\right]} % expectation
\newcommand{\trace}[1] {\mathrm{trace}\LeftP #1 \RightP }

\newcommand{\Vect}  [1] {\mathrm{vec}  \LeftP #1 \RightP } % vectorizing a matrix
\newcommand{\diag} [1] {\mathrm{diag} \LeftP #1 \RightP }
\newcommand{\adiag}[1] {\ma{\Lambda}^{(#1)}} % diagonal channel matrix
\newcommand{\unvec} [3] {\mathrm{unvec}_{#2 \times #3} \LeftP #1 \RightP }% reshape vector to matrix

\newcommand{\modulo }[2] {<#1>_{#2} }% kronecker product

\newcommand{\DFT} [1] {\ma{F}_{#1}} % DFT matrix
\newcommand{\Hdiag} {\adiag{\dft{h}}} % diagonal channel matrix

\newcommand{\U}  [2]{ \ma{U}_{#1,#2} } % Mask matrix
\newcommand{\PI}  [2]{ \ma{\Pi}_{#1,#2} } % permutaiton matrix
\newcommand{\V}  [3]{ \ma{V}^{(#3)}_{#1,#2} } % V matrix
\usepackage{lipsum}
\usepackage{mathtools}
\usepackage{cuted}
\usepackage{amsthm}
\newtheorem{mydef}{Definition}
\newtheorem{mythero}{Lemma}

\begin{document}
\begin{acronym}
	\acro{ZMCSC}{zero-mean circular symmetric complex
	Gaussian }
	\acro{CEq}{channel equalization}
	\acro{OTFS}{Orthogonal Time Frequency Space}
	\acro{FD}{frequency domain}
	\acro{TD}{time domain}
	\acro{OOB}{out-of-band}
	\acro{RRC}{root-raised Cosine}
	\acro{RC}{raised-cosine}
	\acro{ISI}{inter-symbol-interference}
	\acro{ZF}{zero-forcing}
	\acro{MF}{matched filter}
	\acro{SINR}{signal-to-interference-plus-noise ratio}
	\acro{SNR}{signal-to-noise ratio}
	\acro{FIR}{finite impulse repose }
	\acro{DFT}{discrete Fourier transform}
	\acro{OFDM}{orthogonal frequency division multiplexing}
	\acro{GFDM}{generalized frequency division multiplexing}
	\acro{ICI}{inter-carrier-interference}
	\acro{IAI}{inter-antenna-interference}
	\acro{NEF}{noise-enhancement factor}
	\acro{FDE}{frequency domain equalization}
	\acro{SVD}{singular-value decomposition}
	\acro{AWGN}{additive white Gaussian noise}
	\acro{DTFT}{discrete-time Fourier transform}
	\acro{FFT}{fast Fourier transform}
	\acro{SIR}{signal-to-interference ratio}
	\acro{DZT}{discrete Zak transform}
	\acro{MIMO}{multiple-input multiple-output}
	\acro{PAPR}{peak-to-average power ratio}
	\acro{F-OFDM}{filtered OFDM}
	\acro{CP}{cyclic prefix}
	\acro{CS}{cyclic suffix}
	\acro{ZP}{zero padding}
	\acro{IBI}{inter-block-interference}
	\acro{GT}{guard tone}
	\acro{UF-OFDM}{universal-filtered OFDM}
	\acro{FBMC}{filter bank multicarrier}
	\acro{OQAM}{offset quadrature amplitude modulation}
	\acro{FER}{frame error rate}
	\acro{MMSE}{minimum mean square error}
	\acro{IAI}{inter-antenna-interference}
	\acro{MCS}{modulation coding scheme}
	\acro{PSD}{power spectral density}
	\acro{IoT}{Internet of Things}
	\acro{MTC}{machine-type communication}
	\acro{STC}{space-time coding}
	\acro{TR-STC}{time-reversal space-time coding}
	\acro{MRC}{maximum-ratio combiner}
	\acro{LS}{least squares}
	\acro{LMMSE}{linear minimum mean squared error}
	\acro{CIR}{channel impulse response}
	\acro{STO}{symbol time offset}
	\acro{CFO}{carrier frequency offset}
	\acro{UE}{user equipment}
	\acro{FO}{frequency offset}
	\acro{TO}{time offset}
	\acro{BS}{base station}
	\acro{FMT}{filtered multitone }
	\acro{DAC}{digital-to-analogue converter }
	\acro{FO}{frequency offset}
	\acro{TO}{time offset}
	\acro{ISI}{inter-symbol-interference}
	\acro{IUI}{inter-user-interference}
	\acro{IBI}{inter-block-interference}
	\acro{i.i.d.}{independent and identically distributed}
	\acro{SER}{symbol error rate}
	\acro{LTE}{Long Term Evolution}
	% Shahab
	\acro{SISO}{single-input single-output}
	\acro{Rx}{receive}
	\acro{Tx}{transmit}
	\acro{MSE}{mean squared error}
	\acro{IFPI}{interference-free pilot insertion}
	\acro{PDP}{power-delay-profile}
	\acro{ML}{maximum likelihood}
	% nicola
	\acro{5G}{5th generation}
	\acro{4G}{4th generation}
	\acro{NR}{New Radio}
	\acro{eMBB}{enhanced media broadband}
	\acro{URLLC}{ultra-reliable and low-latency communication}
	\acro{mMTC}{massive machine type communication}
	\acro{SDR}{software defined radio}
	\acro{RF}{radio frequency}
	\acro{PHY}{physical layer}
	\acro{MAC}{medium access layer}
	\acro{FPGA}{field programmable gate array}
	\acro{IDFT}{inverse discrete Fourier transform}
	\acro{DRAM}{dynamic random access memory}
	\acro{BRAM}{block RAM}
	\acro{FIFO}{first in first out}
	\acro{D/A}{digital to analog}
	\acro{EVA} {extended vehicular A channel model} 
	\acro{OTFS}{Orthogonal time frequency space modulation}
	\acro{SFFT}{symplectic finite Fourier transform}
	
	% Henry
	\acro{ACLR}{adjacent channel leakage rejection}
	\acro{ADC}{analog-to-digital converter}
	\acro{AGC}{automatic gain control}
	\acro{CEP}{channel estimation preamble}
	\acro{DPD}{digital pre-distortion}
	\acro{PA}{power amplifier}
	\acro{LTV}{linear time-variant}
	
	%MA
	\acro{NMSE}{normalized mean-squared error}
	\acro{PRB}{physical resource block}
	\acro{BER}{bit error rate}
	\acro{FER}{frame error rate}
	\acro{DL}{downlink}
	\acro{UL}{uplink}
	\acro{FO}{frequency offset}
	\acro{TO}{time offset}
	\acro{MA}{multiple access}
	
	\acro{INI}{inter-numerology-interference}
	\acro{PCCC}{parallel concatenated convolutional code}
	\acro{CCDF}{complementary cumulative distribution function}
	\acro{SC}{single carrier}
	\acro{FDMA}{frequency division multiple access}
	\acro{IP}{intellectual property}
	\acro{CM}{complex multiplication}
	\acro{DSP}{digital signal processor }
	\acro{LUT}	{lookup table}
	\acro{RAM}	{random-access memmory}
	\acro{RW}	{read-and-write}
	\acro{R/W}	{read-or-write}
	\acro{MCM}{multicarrier modulation}
	\acro{PAPR}{peak-to-average power ratio}
	\acro{FDMA}{frequency division multiple access}
	\acro{GFDMA}{generalized frequency division multiple access}
\end{acronym}

\title{Practical GFDM-based Linear Receivers	
\thanks{The work presented in this paper has been performed in the framework of the ORCA project [https://www.orca-project.eu/].This project has received funding from the European Union's Horizon 2020 research and innovation programme under grant agreement No 732174.}}
\begin{comment}
\author{
\IEEEauthorblockN{Ahmad Nimr, Marwa Chafii,    Gerhard Fettweis}
\IEEEauthorblockA{ Vodafone Chair Mobile Communication Systems, Technische Universit\"{a}t Dresden, Germany}
\IEEEauthorblockA{\small\texttt{\{first name.last name\}@ifn.et.tu-dresden.de}}
}
\begin{comment}
\author{
	\IEEEauthorblockN{Author 1, Author 2, Author 3,  Author 4}\\
	%\IEEEauthorblockA{ Vodafone Chair Mobile Communication Systems, Technische Universit\"{a}t Dresden, Germany}\\
	%\IEEEauthorblockA{\small\texttt{\{first name.last name\}}}\IEEEauthorblockA{\small\texttt{@ifn.et.tu-dresden.de}}
}
\end{comment}
\author{
	\IEEEauthorblockN{Ahmad Nimr\IEEEauthorrefmark{1}, Marwa Chafii\IEEEauthorrefmark{2},  Gerhard Fettweis\IEEEauthorrefmark{1}}
	\IEEEauthorblockA{\IEEEauthorrefmark{1}Vodafone Chair Mobile Communication Systems, Technische Universit\"{a}t Dresden, Germany}
	\IEEEauthorblockA{\IEEEauthorrefmark{2} ETIS UMR 8051, Université Paris Seine, Université de Cergy-Pontoise, ENSEA, CNRS, France}
	\IEEEauthorblockA{\small\texttt{ahmad.nimr@ifn.et.tu-dresden.de, marwa.chafii@ensea.fr, gerhard.fettweis@tu-dresden.de}}
}

%\author{GFDM }
\maketitle
\IEEEpeerreviewmaketitle
\begin{abstract}
The conventional receiver designs of \ac{GFDM} consider a large scale \ac{MIMO} system with a block circular matrix of combined  channel and modulation. Exploiting this structure, several approaches have been proposed for low complexity joint \ac{LMMSE} receiver. However, the joint design is complicated and inappropriate for hardware implementation. In this paper, we define the concept of \ac{GFDM}-based linear receivers, which first performs  \ac{CEq} and afterwards the equalized signal is processed with \ac{GFDM} demodulator.
We show that the optimal joint \ac{LMMSE} receiver is equivalent to a \ac{GFDM}-based one, that applies \ac{LMMSE}-\ac{CEq} and \acl{ZF} demodulation. For  orthogonal modulation, the optimal \ac{LMMSE} receiver has an implementation-friendly structure. For the non-orthogonal case, we propose two practical designs that approach the performance of the joint \ac{LMMSE}. Finally, we analytically prove that \ac{GFDM}-based receivers achieve equal \acl{SINR} per subsymbols within the same subcarrier.
\end{abstract}
%%% Local Variables:
%%% mode: latex
%%% TeX-master: "FilterDesign"
%%% End:
\acresetall
\section{Introduction}\label{sec:introduction}
In its early proposal, \ac{GFDM} \cite{GFDM} has been suggested as an alternative  to \ac{OFDM}. Recently, \ac{GFDM} has been extended to a  multicarrier framework, that is able to process most of the state of the art waveforms and allows  the design of new  waveforms \cite{danneberg2018universal}. The well-defined structure of \ac{GFDM} enables a feasible real-time modem implementation on hardware. For instance the approaches in \cite{danneberg2015flexible} and \cite{FDE} provide implementations in the \ac{TD} and \ac{FD}, respectively. However, the low complexity receiver approaches in fading channels practically consider \ac{ZF}-\ac{CEq}. One of the main objectives of conventional \ac{GFDM} is providing low \ac{OOB}. This requires a design of the prototype pulse with subcarrier overlapping leading to a non-orthogonal modulation \cite{nimr2017study}. As a consequence, tackling the self \ac{ISI} and \ac{ICI} at the receiver becomes  challenging  in the design of the \ac{GFDM} receiver. By exploiting the sparse representation of the \ac{FD} pulse shape, an interference cancellation approach is proposed in \cite{intefrence_cancellation}. In that work, after \ac{ZF}-\ac{CEq}, and \ac{MF} demodulation, the \ac{ICI} in a subcarrier is canceled based on the hard decision estimation of the symbols from the adjacent subcarriers.
Joint \ac{LMMSE} receiver has been widely studied considering a large scale \ac{MIMO} matrix. The computation of the \ac{LMMSE} filter is complex, especially for  hardware implementation. By exploiting the block diagonal structure of the equivalent \ac{MIMO} channel, the algorithm  in \cite{MMSE_low_max} reduces the complexity of computing the \ac{LMMSE} filter from a cubic to square order. However, the receiver  needs to compute the inverse of  smaller-scale matrices. The works in \cite{MMSE_chen} and \cite{MMSE_Tiwari} reproduce the same results by means of the decomposition of the modulation matrix. A special low-complexity order is given in \cite{MMSE_chen} for orthogonal modulation matrix. Nevertheless, the latter approaches focus more on the theoretical analysis of the performance  without the consideration of hardware implementation.  The joint \ac{LMMSE} performance in terms of  uncoded \ac{BER} is studied in \cite{MMSE_BER} via a closed-form approximation using  the achieved \ac{SINR} per symbol. However, a low-complexity computation of the \ac{SINR} is missing in that work.

In this paper, we aim at a practical implementation of linear \ac{GFDM} receivers  via  decoupled \ac{CEq} and demodulation. We refer to this type of receivers  as \emph{\ac{GFDM}-based} receivers. By means of two-dimensional representation, we show that the received signal in frequency selective channels consists of $M$ parallel uncorrelated small-scale signals. Accordingly, the \ac{GFDM}-based receiver
as well as the \acp{SINR} of the demodulated symbols are analytically computed. We show that \ac{GFDM}-based receivers achieve equal-\ac{SINR} per subsymbols within the same subcarrier. The optimal joint \ac{LMMSE} is equivalent to \ac{LMMSE}-\ac{CEq} and \ac{ZF} demodulation. For an orthogonal modulation matrix, the \ac{LMMSE}-\ac{CEq} requires to compute the inverse of a diagonal matrix. In the non-orthogonal case, we show that the \ac{CEq} can be performed on small-scale parallel signals. Moreover, a low-complexity practical approximation of \ac{LMMSE}-\ac{CEq} is derived. Furthermore, we investigate  the \ac{LMMSE} demodulation after \ac{ZF}-\ac{CEq}. Both approximations approach the performance of the optimal joint \ac{LMMSE}, while allowing a  practical implementation

The remainder of the paper is organized as follows: Section \ref{sec:system model} introduces the system model and an overview of \ac{GFDM} \ac{FD} modem  structure. Section \ref{sec:SISO receiver} is dedicated for the design of \ac{GFDM}-based receivers. In Section \ref{sec:RX alternative}, we focus on the parallel system model which is used to analytically drive the receivers and the \ac{SINR} expressions. Section \ref{sec: evaluation} provides numerical results.  Finally, Section \ref{sec:conclusions} concludes the paper.
\section{System Model}\label{sec:system model}
We consider a \ac{GFDM} system with $K$ subcarriers, $M$ subsymbols, and a prototype pulse $g[n]$. The \ac{GFDM} block \small $\ma{x}\in \compl ^{N\times 1 },~ N= KM$\normalsize, is given by \cite{GFDM}
\begin{equation}
\small
\IndexV{\ma{x}}{n} = \sum\limits_{k=0}^{K-1}\sum\limits_{m = 0}^{M-1}\IndexM{\ma{D}}{k}{m} g[\modulo {n -mK}{N}] e^{j2\pi\frac{k}{K}n}, \label{eq:basic equation TD}
\end{equation}
where \small$\modulo {\cdot}{N}$ \normalsize is the modulo-$N$ operator and  \small $\ma{D} \in \compl^{K\times M}$ \normalsize is the input data matrix. The data symbol \small $d_{k,m} = \IndexM{\ma{D}}{k}{m}$ \normalsize is transmitted on the  $m$-th subsymbol of the $k$-th subcarrier. The \ac{GFDM} block can be expressed in a matrix notation as
\begin{equation}
\small
\ma{x} = \ma{A}\ma{d},~\IndexM{\ma{A}}{n}{k+mK} = g[\modulo {n-mK}{N}]e^{j2\pi\frac{kn}{K}}, \label{eq:matrix representation}
\end{equation}
where \small $\ma{d} = \Vect{\ma{D}}$\normalsize.
A frame of \ac{GFDM} blocks is transmitted over block fading wireless multipath channel with impulse response $h[l]$. To enable \ac{FD} equalization, a \ac{CP} longer than the channel delay spread is appended to the beginning of each \ac{GFDM} block. After removing the \ac{CP}, we get the received block \small${\ma{y}} = \ma{H} {\ma{A}}\ma{d} + {\ma{v}}$\normalsize, 
%\begin{equation}
%\small
%\begin{split}
%{\ma{y}} &= \ma{H} {\ma{A}}\ma{d} + {\ma{v}},\label{eq:received GFDM signal TD}
%\end{split}
%\end{equation}
where \small$\ma{H}\in \compl^{N\times N}$ \normalsize is the circular channel matrix, \small$\IndexM{\ma{H}}{n}{q} = h[\modulo {n-q}{N}]$\normalsize. The  \ac{AWGN} vector with variance $\sigma^2$ is denoted as $\ma{v}$. 
By applying $N$-\ac{DFT}, the \ac{FD} received block is written as 
\begin{equation}
\small
\begin{split}
\dft{\ma{y}} &= \Hdiag \dft{\ma{A}}\ma{d} + \dft{\ma{v}},\label{eq:received GFDM signal }
\end{split}
\end{equation}
where
 \small$\Hdiag = \diag{\{\IndexV{\dft{\ma{h}}}{n}\}_{n=0}^{N-1}}$ \normalsize is the equivalent \ac{FD} diagonal channel matrix,  \small$\dft{\ma{h}} = N\mbox{-DFT}\{h[l]\}$\normalsize.
The notation \small$\dft{\ma{X}} = \DFT{N}\ma{X}$ \normalsize denotes the $N$-\ac{DFT} of the columns of $\ma{X}$. 
\subsection{Joint receiver}
This approach considers the general  \ac{MIMO} system \cite{MMSE_Tiwari} 
\begin{equation}
\small
\dft{y} = \ma{H}^{(\text{eff})} \ma{d} + \dft{\ma{v}}, ~\ma{H}^{(\text{eff})} = \Hdiag \dft{\ma{A}}.
\end{equation}
All the approaches of \ac{MIMO} receiver can be applied. The structure of $\dft{ \ma{A}}$ can be exploited for low complexity computation. For instance, 
the joint \ac{ZF} (\small $\ma{H}^{(\text{eff})-1}$ \normalsize) is decoupled into \ac{ZF}-\ac{CEq} (\small ${\Hdiag}^{-1}$ \normalsize) and \ac{ZF}-\ac{GFDM}-demodulator (\small $\dft{\ma{A}}^{-1}$\normalsize). Moreover, assuming uncorrelated data, i.e. \small$\ma{R}_d = \Ex{\ma{d}\ma{d}^{H}} = E_s\ma{I}_{N}$\normalsize, where $E_s$ is the average symbol power, and \small $\ma{R}_{\dft{v}} = \Ex{\dft{\ma{v}}\dft{\ma{v}}^H} = \dft{\sigma}^2\ma{I}_N$, $\dft{\sigma}^2 = N{\sigma}^2$\normalsize,
the \ac{LMMSE} receiver filter  is given by
\begin{equation}
\small
\begin{split}
\ma{W}^H &= \ma{H}^{(\text{eff}) H}\left(\ma{H}^{(\text{eff})}\ma{H}^{(\text{eff})H}+\frac{\dft{\sigma}^2}{E_s}\ma{I}_N\right)^{-1}.
\end{split}
\end{equation}
In the case of orthogonal modulation matrix, i.e. \small$\ma{A}\ma{A}^H = \ma{I}_N\Rightarrow \dft{\ma{A}}\dft{\ma{A}}^H = N\ma{I}_N$\normalsize, the joint \ac{LMMSE} is reduced to 
\begin{equation}
\small
\begin{split}
\ma{W}^H &= \frac{1}{N}\dft{\ma{A}}^H\underbrace{{\Hdiag}^{H}\left(\Hdiag{\Hdiag}^H+\frac{\dft{\sigma}^2}{E_s}\ma{I}_N\right)^{-1}}_{\Hdiag_{\text{eq}}}.
\end{split}
\end{equation}
This can be computed by first performing \ac{CEq} with the matrix $\Hdiag_{\text{eq}}$ and then \ac{MF}-\ac{GFDM}-demodulation. In this case, the \ac{LMMSE}  implementation  is feasible. On the contrary, when $\ma{A}$ is non-orthogonal, the hardware realization of the joint \ac{LMMSE} is not affordable.
\subsection{Frequency-domain modem}
\begin{figure}[h]
	\centering
	\includegraphics[width=.95\linewidth]{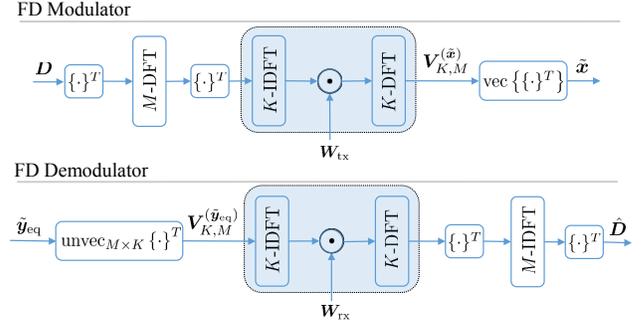}
	\caption{\ac{GFDM} \ac{FD} modem, the highlighted box represents \eqref{eq:advanced convolution FD}.}\label{fig:Advanced_FD}
\end{figure}
 The structure of the \ac{FD} modulation matrix  $\dft{\ma{A}}$ is  derived from the \ac{FD} block representation
\begin{equation}
\small
\IndexV{\dft{\ma{x}}}{n} = \sum\limits_{m=0}^{M-1}\sum\limits_{k = 0}^{K-1}\IndexM{\ma{D}}{k}{m} \tilde{g}[\modulo {n -kM}{N}] e^{-j2\pi\frac{m}{M}n}. \label{eq:basic equation FD}
\end{equation}
Here, $\tilde{g}[n]$ is the \ac{FD} prototype pulse. By reformulating \eqref{eq:basic equation FD}  using two indexes \small$p = 0,\cdots, M-1$ \normalsize and \small$q = 0,\cdots, K-1$\normalsize, with \small $n = p + qM$\normalsize, we get
\small 
\begin{equation*}
\begin{split}
\IndexV{\dft{\ma{x}}}{p+qM}
= \sum\limits_{k=0}^{K-1} \dft{g}[<p+[q-k]M>_N]\sum\limits_{m=0}^{M-1} \IndexM{\ma{D}}{k}{m} e^{-j2\pi\frac{mp}{M} }.
\end{split}
\end{equation*}
\normalsize
\begin{figure}[t]
	\centering
	\includegraphics[width=1\linewidth]{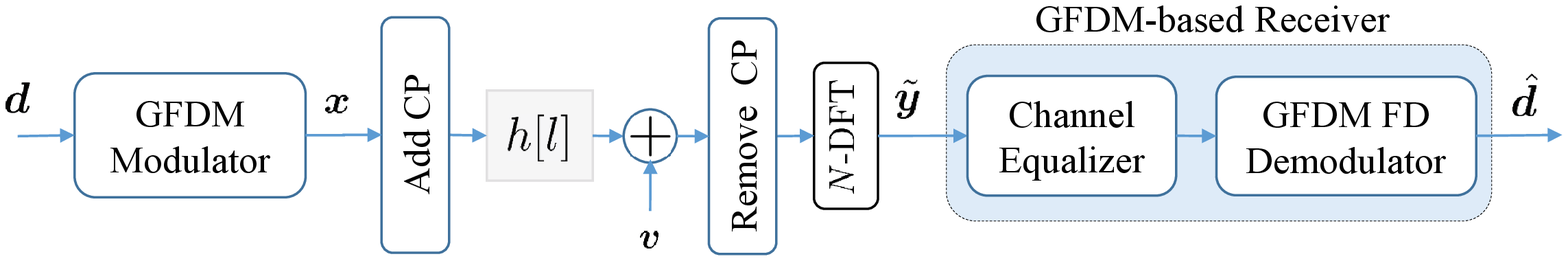}
	\caption{\ac{GFDM} baseband system model.}\label{fig:GFDM_based_receiver}
\end{figure}
The matrix notation \small$\V{K}{M}{\dft{\mav{x}}} \in \compl^{K\times M}$ \normalsize  is defined such that
\begin{equation}
\small
\IndexM{\V{K}{M}{\dft{\mav{x}}}}{q}{p} = \IndexV{\dft{\ma{x}}}{p+qM} \Leftrightarrow
\V{K}{M}{\dft{\mav{x}}} = \unvec{\dft{\mav{x}}}{M}{K} ^T. \label{eq:polyphase time}
\end{equation}
\small
\begin{equation*}
\begin{split}
\mbox{Thus, }\small
\IndexM{\V{K}{M}{\dft{\ma{x}}}}{q}{p}
= \sum\limits_{k=0}^{K-1}\IndexM{\V{K}{M}{\dft{\ma{g}}}}{<q-k>_K}{p}\IndexM{\ma{D}\DFT{M}}{k}{p}.
\end{split} 
\end{equation*}
\normalsize
This defines a  circular convolution between the $p$-th column of \small$\V{K}{M}{\dft{\ma{g}}}$ \normalsize and the $p$-th column of \small$\ma{D}\DFT{M}$\normalsize. Accordingly,
\begin{equation}
\small
\begin{split}
\V{K}{M}{\dft{\ma{x}}} = \V{K}{M}{\dft{\ma{g}}}\circledast_1[\ma{D}\DFT{M}],
\end{split}
\label{eq:advanced convolution FD}
\end{equation}
where $\circledast_1$ denotes the circular convolution with respect to the first dimension, i.e. the columns.  
This circular convolution can be expressed using $K$-I\ac{DFT} as
\small
\begin{equation*}
\small
\begin{split}
\IndexM{\DFT{K}^H\V{K}{M}{\dft{\ma{x}}}}{q}{p} =
\IndexM{\DFT{K}^H\V{K}{M}{\dft{\ma{g}}}}{q}{p} \IndexM{\DFT{K}^H\ma{D}\DFT{M}}{q}{p}.
\end{split}
\end{equation*}
\normalsize
\begin{equation}
\mbox{Therefore, }
\small
{\V{K}{M}{\tilde{\ma{x}}}} =
\DFT{K}\left(\ma{W}_{\text{tx}}\odot\left[\frac{1}{K}\DFT{K}^H{\ma{D}\DFT{M}}\right]\right). \label{eq:advanced ZAK FD}
\end{equation}
Here, $\odot$ denotes the element-wise multiplication operator, and \small $\ma{W}_{\text{tx}}$ \normalsize is the modulator window, which is derived from $\dft{\ma{g}}$ as
\begin{equation}
\small \ma{W}_{\text{tx}}= \DFT{K}^H\V{K}{M}{\dft{\ma{g}}} \in \compl^{K\times M}\label{eq:zak frequency}.
\end{equation}
Fig.~\ref{fig:Advanced_FD} illustrates the block diagram of the \ac{FD} modem. The highlighted box corresponds to the convolution  \eqref{eq:advanced convolution FD}.
The \ac{FD} demodulator performs the inverse operations on the \ac{FD} equalized  block $\dft{\ma{y}}_{\text{eq}}$ using a receiver window \small$\ma{W}_{\text{rx}}$\normalsize, 
\begin{equation}
\small
\hat{\ma{D}} = \frac{1}{M}\DFT{K}\left(\ma{W}_{\text{rx} }\odot\left[\frac{1}{K}\DFT{K}^H{\V{K}{M}{\dft{\ma{y}}_{\text{eq}}}}\right]\right)\DFT{M}^H.
\end{equation}
\subsection{GFDM matrix structure }
The structure of $\dft{\ma{A}}$ can be revealed by the vectorization of \eqref{eq:advanced ZAK FD}, where \small$ \dft{\ma{x}} = \Vect{{\V{K}{M}{\dft{\ma{x}}}}^T} = \dft{\ma{A}}\ma{d}$\normalsize. As a result
\begin{equation}
\small
\dft {\ma{A}}=\underbrace{\PI{K}{M}\U{M}{K}}_{\ma{V}_f}{\ma{\Lambda}}^{(\text{tx})}\underbrace{\U{M}{K}^H\PI{M}{K}\U{K}{M}\PI{K}{M}}_{\ma{U}_t}.\label{eq:FD decompostion}
\end{equation}
Here \small$\PI{Q}{P} $ \normalsize is the commutative matrix of size \small $QP\times QP$ \normalsize defined such that
for  a matrix \small $\ma{X}\in \compl^{Q\times P}$, $\Vect{\ma{X}^T} = \PI{Q}{P}\Vect{\ma{X}}$, $\U{P}{Q}$ \normalsize is unitary matrix given by \small $\U{P}{Q} = \frac{1}{\sqrt{Q}}\ma{I}_P\otimes\DFT{Q}$\normalsize, and \small $\ma{\Lambda}^{(\text{tx})}$ \normalsize is a diagonal matrix given by
\begin{equation}
\small
\ma{\Lambda}^{(\text{tx})} = \sqrt{M} \diag{\Vect{\ma{W}_{\text{tx}}}}.
\end{equation}
Note that $\ma{V}_f$ and $\ma{U}_t$ are unitary matrices. Hence, we define
\begin{mydef}
	An \ac{FD} \ac{GFDM} matrix of $K$ subcarriers and $M$ subsymbols is a square matrix of size $N\times N$ that can be decomposed according to  \eqref{eq:FD decompostion}.
\end{mydef}
\noindent From the demodulator structure, the \ac{FD} demodulator matrix \small$\dft{\ma{B}}\in \compl^{N\times N}$\normalsize, where \small$\hat{\ma{d}} = \dft{\ma{B}}^H\dft{\ma{y}}_{\text{eq}}$ \normalsize is a \ac{GFDM} matrix given by
\begin{equation}
\small
\dft {\ma{B}}=\underbrace{\PI{K}{M}\U{M}{K}}_{\ma{V}_f}{\ma{\Lambda}}^{(\text{rx})H}\underbrace{\U{M}{K}^H\PI{M}{K}\U{K}{M}\PI{K}{M}}_{\ma{U}_t}.
\end{equation}
\begin{equation*}
\small
\begin{split}
\mbox{Here,~ }\small \ma{\Lambda}^{(\text{rx})} = \frac{1}{\sqrt{M}}\diag{\Vect{\ma{W}_{\text{rx}}}}.
\end{split}
\end{equation*}
The design of the demodulator is achieved by computing the diagonal matrix $\ma{\Lambda}^{(\text{rx})}$ or equivalently the window $\ma{W}_{\text{rx}}$.
\section{GFDM-based Receiver}\label{sec:SISO receiver}
In a realistic implementation of  \ac{GFDM} receiver, a \ac{CEq} precedes the demodulation. Usually a simple \ac{ZF}-\ac{CEq} is applied \cite{danneberg2015flexible}. By considering  independent designs of the \ac{CEq} and demodulation, we formulate the following definition:
\begin{mydef}
	A \ac{GFDM}-based receiver is a receiver that first, performs \ac{CEq},  and then, the equalized signal is demodulated with a \ac{GFDM} demodulator.
\end{mydef}
\noindent This definition can be applied for \ac{TD} or \ac{FD} processing. In this work, we focus on \ac{FD} \ac{GFDM}-based receiver. Namely, an \ac{FD} channel equalizer and \ac{FD} demodulator as illustrated in  Fig.~\ref{fig:GFDM_based_receiver}.
\subsection{Relation to joint \ac{LMMSE} receiver }
The relation between a \ac{GFDM}-based receiver and the joint \ac{LMMSE} receiver is summarized by the following lemma:
\begin{mythero}
	If  \small$\Hdiag, \dft{\ma{A}}$ \normalsize are invertible, and the data and noise are uncorrelated, such that  \small$\ma{R}_d = E_s\ma{I}_N$\normalsize, and \small $\ma{R}_{\dft{v}}  = \dft{\sigma}^2\ma{I}_N$\normalsize, the \ac{LMMSE} receiver can be computed in two ways: 
	\begin{enumerate}
		\item \ac{LMMSE}-\ac{CEq} on the linear model\\\small${\dft{\ma{y}}_{\text{eq}} = \Hdiag \dft{\ma{x}} + \ma{v},~ \ma{R}_{\dft{x}} = E_s\dft{\ma{A}}\dft{\ma{A}}^H}$\normalsize\\
		followed by \ac{ZF} demodulation, $ \hat{\ma{d}} = \dft{\ma{A}}^{{-1}}\dft{\ma{y}}_{\text{eq}}$.
		\item  \ac{ZF}-\ac{CEq}, i.e. $\dft{\ma{y}}_{\text{eq}} = {\Hdiag}^{-1}\ma{y}$, followed by \ac{LMMSE} demodulation on\\ 
\small	${	\dft{\ma{y}}_{\text{eq}}=  \dft{\ma{A}}\ma{d} + \bar{\ma{v}},~\ma{R}_{\bar{v}} = \dft{\sigma}^2\left[\Hdiag {\Hdiag}^{H}\right]^{-1}}$\normalsize.
	\end{enumerate}
\end{mythero}\label{lemma 1}
\noindent The proof is given in Appendix \ref{sec: MMSE appindx}. The first method,  represents a  decoupled \ac{LMMSE}-\ac{CEq},
\begin{equation}
\small
\dft{\ma{H}}_{\text{LMMSE}}^H = \left({\Hdiag} ^H\Hdiag  + \frac{\dft{\sigma}^2}{E_s}\left[\dft{\ma{A}}\dft{\ma{A}}^H\right]^{-1}\right)^{-1}{\Hdiag} ^H,
\end{equation}
followed by a \ac{ZF}-\ac{GFDM}-demodulation with the demodulator window \small$\IndexM{\ma{W}_{\text{rx}}}{k}{m} = \left[\IndexM{\ma{W}_{\text{tx}}}{k}{m}\right]^{-1}$\normalsize.
This case follows the definition of a \ac{GFDM}-based receiver. If $\ma{R}_{\dft{x}}$ is diagonal, e.g. when  $\ma{A}$ is orthogonal, the \ac{LMMSE}-\ac{CEq} is reduced to the computation of the inverse of a diagonal matrix, which is simple for realization. Otherwise, the main complexity is inherited from the computation of the inverse \small $ {\left({\Hdiag} ^H\Hdiag  + \frac{\dft{\sigma}^2}{E_s}\left[\dft{\ma{A}}\dft{\ma{A}}^H\right]^{-1}\right)}$\normalsize. However, a reduced complexity can be achieved using the decomposition of \eqref{eq:FD decompostion}, where \small $\left[\dft{\ma{A}}\dft{\ma{A}}^H\right]^{-1} = \ma{V}_f\left[\ma{\Lambda}^{(\text{tx})}\ma{\Lambda}^{(\text{tx})H}\right]^{-1} \ma{V}_f^H$\normalsize. This allows the computation of \small$\dft{\ma{H}}_{\text{LMMSE}}^H ~$ \normalsize using the inverse of $M$ matrices each of size $K\times K$ as derived in \cite{MMSE_low_max}. We provide a simplified derivation in Section \ref{sec:  RX alternative equalization}.
In the second approach, the \ac{LMMSE} demodulation following \ac{ZF}-\ac{CEq} is given by
\begin{equation}
\small
\begin{split}
\ma{B}_{\text{MMSE}}^H &= \dft{\ma{A}}^H\left(\dft{\ma{A}}\dft{\ma{A}}^H+\frac{\dft{\sigma}^2}{E_s}\ma{R}_{\bar{v}}\right)^{-1}\\
&= \ma{U}_t^H\underbrace{{\ma{\Lambda}}^{(\text{tx} )H} \left({\ma{\Lambda}}^{(\text{tx})}{\ma{\Lambda}}^{(\text{tx} )H} + \ma{V}_f^H\frac{\ma{R}_{\bar{v}}}{E_s} \ma{V}_f\right)^{-1}}_{\ma{\Gamma^{(\text{rx})}}}\ma{V}_f^H. \label{eq: demodulator LMMSE exact}
\end{split}
\end{equation}
If the matrix $\ma{\Gamma^{(\text{rx})}}$ is diagonal, e.g.  \ac{AWGN} \cite{FDE} channel, then $\ma{B}_{\text{MMSE}}$ becomes a \ac{GFDM} matrix. Otherwise, $\ma{B}_{\text{MMSE}}^H$  cannot be implemented with \ac{GFDM}-demodulator. Nevertheless, the demodulator can be designed with \ac{LMMSE}  under the constraint of \ac{GFDM} matrix, as discussed in Section \ref{sec: RX alternative LMMSE demod}.
\section{GFDM parallel signal model }\label{sec:RX alternative}
Using \eqref{eq:advanced convolution FD} and \eqref{eq:polyphase time}, the received signal can be represented as
\begin{equation}
\small
\begin{split}
\V{K}{M}{\dft{ \ma{y}}} &= \V{K}{M}{\dft{ \ma{h}}}\odot \left(\V{K}{M}{\dft{\ma{g}}}\circledast_1[\ma{D}\DFT{M}]\right)+ \V{K}{M}{\dft{ \ma{v}}}. \label{eq: received signal conv}
\end{split}
\end{equation}
 \begin{figure}[h]
	\centering
	\includegraphics[width=1\linewidth]{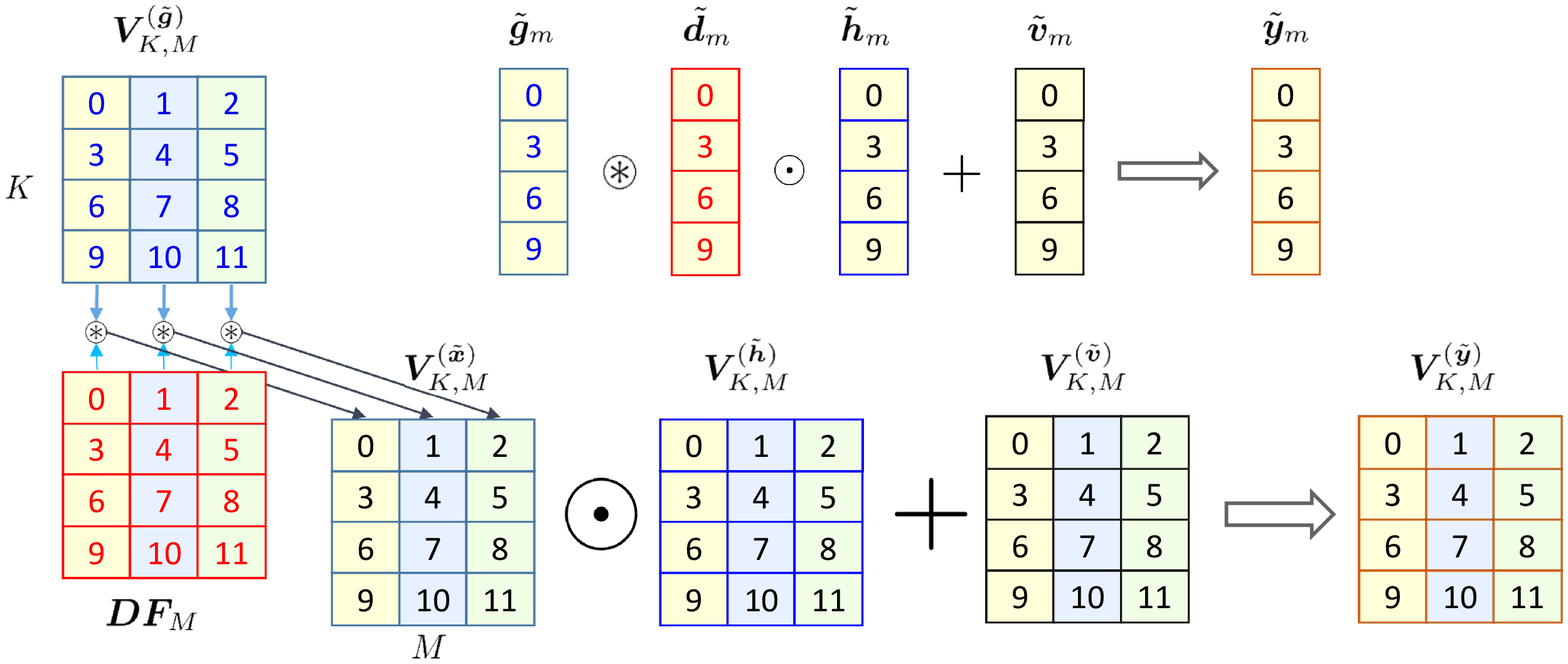}
	\caption{ Parallel signal model.}\label{fig:SISO_receiver}
\end{figure}
As illustrated in Fig.~\ref{fig:SISO_receiver}, the $m$-th column, \small $\dft{\ma{y}}_m = \IndexM{\V{K}{M}{\dft{ \ma{y}}}}{:}{m}$\normalsize, can be written in the form 
\begin{equation}
\small
\begin{split}
\dft{\ma{y}}_m &= \Hdiag_m \dft{\ma{A}}_m \dft{\ma{d}}_m + \dft{\ma{v}}_m,  \dft{\ma{v}}_m = \IndexM{\V{K}{M}{\dft{ \ma{v}}}}{:}{m}\in \compl^{K\times 1},
\label{eq:parallel signal}
\end{split}
\end{equation}
where, \small $\dft{\ma{d}}_m = \IndexM{\ma{D}\DFT{M}}{:}{m} ,~\Hdiag_m = \diag{\IndexM{\V{K}{M}{\dft{ \ma{h}}}}{:}{m}}$\normalsize, and  $\dft{\ma{A}}_m \in \compl^{K\times K}$ is a circular matrix generated from the column vector $\dft{\ma{g}}_m =   \IndexM{\V{K}{M}{\dft{ \ma{g}}}}{:}{m}
$. It can be expressed by means of  $K$-\ac{DFT} and the modulator window \eqref{eq:zak frequency} as
\begin{equation}
\small
\dft{\ma{A}}_m = \frac{1}{K}\DFT{K} \ma{\Lambda}^{(\text{tx})}_m\DFT{K}^H,~ \ma{\Lambda}^{(\text{tx})}_m = \diag{\IndexM{\ma{W}_{\text{tx}}}{:}{m}}.
\end{equation}
Considering full allocation, uncorrelated data, and uncorrelated noise, then \small $\Ex{\dft{\ma{d}}_m \dft{\ma{d}}_m ^H} = ME_s\ma{I}_K$ \normalsize and 
\small $\Ex{\dft{\ma{d}}_m \dft{\ma{d}}_p^H} = \ma{0}_K, p\neq m$\normalsize. Therefore, \small $\Ex{\dft{\ma{y}}_m \dft{\ma{y}}_p^H} = \ma{0}_K, p\neq m$\normalsize. This means that the received signal can be decoupled into $M$ parallel uncorrelated signals, each has the size of $K$ samples.
\subsection{LMMSE-\ac{CEq}}\label{sec:  RX alternative equalization}
The \ac{LMMSE} \ac{CEq} can be performed on the signal \eqref{eq:parallel signal}. Let \small $\dft{\ma{H}}^H_{m,\text{eq}}$ \normalsize be the channel equalizer given by
\begin{equation}
\small
\dft{\ma{H}}^H_{m,\text{eq}} = \left({\Hdiag_m}^H\Hdiag_m + {\dft{\sigma}^2}\ma{R}_{\bar{d},m}^{-1}\right)^{-1}{\Hdiag_m}^H,\label{eq:CEq lmmase full}
\end{equation}
where  \small$\ma{R}_{\bar{d},m} = ME_s\dft{\ma{A}}_m \dft{\ma{A}}_m^H = \frac{ME_s}{K}\DFT{K} \ma{\Lambda}^{(\text{tx})}_m\ma{\Lambda}^{(\text{tx})H}_m\DFT{K}^H$\normalsize.
If the elements of $\ma{\Lambda}^{(\text{tx})}_m$ are of equal amplitude then $\ma{R}_{\bar{d},m}$ is diagonal and thus  \small $\dft{\ma{H}}^H_{m,\text{eq}}$  \normalsize is diagonal. For instance, in conventional \ac{GFDM} with two subcarrier overlap, depending on the roll-off factor, there is several indexes where  $\ma{\Lambda}^{(\text{tx})}_m$ satisfies the equal amplitude condition \cite{nimr2017optimal}.  Otherwise, a practical \ac{LMMSE} under diagonal matrix constraint is obtained using \small$\diag{\ma{R}_{\bar{d},m}} = \frac{ME_s}{K}P_m^{(\text{tx})} \ma{I}_K$\normalsize, where
\begin{equation}
\small
P_m^{(\text{tx})} =\trace{\ma{\Lambda}^{(\text{tx})}_m\ma{\Lambda}^{(\text{tx})H}_m}.
\end{equation}
\begin{equation}
\mbox{Then, }\small
\ma{\Lambda}^{(\dft{\ma{h}})}_{m,\text{eq}} = \left({\Hdiag_m}^H\Hdiag_m + \frac{K\dft{\sigma}^2}{ME_sP_m^{(\text{tx})} }\ma{I}_K\right)^{-1}{\Hdiag_m}^H.\label{eq:CEq lmmase diag}
\end{equation}
The implementation of this receiver requires the knowledge of  $P_m$, which can be acquired in advance. The inverse is realized with a real-valued reciprocal block. In addition, one block to compute the absolute value and a complex multiplier are required to compute the \ac{CEq}. 
\subsection{LMMSE-\ac{GFDM}-demodulation}\label{sec: RX alternative LMMSE demod}
After applying \ac{ZF}-\ac{CEq} on \eqref{eq:parallel signal}, we get 
\begin{equation}
	\small
\dft{\ma{y}}_{\text{eq}, m} =  \frac{1}{K}\DFT{K} \ma{\Lambda}^{(\text{tx})}_m\DFT{K}^H \dft{\ma{d}}_m + {\Hdiag_m}^{-1}\dft{\ma{v}}_m.
\end{equation}
The next step of the modulator is to apply $K$-I\ac{DFT}, 
\begin{equation}
\small
\frac{1}{K}\DFT{K}^H\dft{\ma{y}}_{\text{eq}, m} =  \frac{1}{K} \ma{\Lambda}^{(\text{tx})}_m\DFT{K}^H \dft{\ma{d}}_m + \frac{1}{K}\DFT{K}^H{\Hdiag_m}^{-1}\dft{\ma{v}}_m.
\end{equation}
The \ac{LMMSE} window can then be computed as
\begin{equation}
\small
\ma{\Gamma}^{(\text{rx})}_m =  \ma{\Lambda}^{(\text{tx})H}_{m}\left(\ma{\Lambda}^{(\text{tx})}_{m}\ma{\Lambda}^{(\text{tx})H}_{m} + \frac{K}{ME_s}\ma{R}_{v_s,m} \right)^{-1},
\end{equation}
\begin{equation*}
\small
\begin{split}
\mbox{where~}
\small
\ma{R}_{d_s,m} = \frac{ ME_s}{K}\ma{I}_K,~\ma{R}_{v_s,m} = \frac{\dft{\sigma}^2}{K^2} \DFT{K}^H\left[{\Hdiag_m}{\Hdiag_m}^H\right]^{-1}\DFT{K}.
\end{split}
\end{equation*} 
In general $\ma{\Gamma}^{(\text{rx})}_m$ is not diagonal. To allow \ac{GFDM} demodulation, we consider the constraint of diagonal matrix, which is achieved by using \small$\diag{\ma{R}_{v_s,m}} = \frac{\dft{\sigma}^2}{K^2}\Omega_m^{(\dft{h})} \ma{I}_K$\normalsize, where 
\begin{equation}
\small
\Omega_m^{(\dft{h})} = \trace{\left[{\Hdiag_m}{\Hdiag_m}^H\right]^{-1}}. 
\end{equation}
\begin{equation}
\mbox{Thus, }\small
\ma{\Lambda}^{(\text{rx})}_{m} = \ma{\Lambda}^{(\text{tx})H}_{m}\left(\ma{\Lambda}^{(\text{tx})}_{m}\ma{\Lambda}^{(\text{tx})H}_{m} + \frac{\dft{\sigma}^2 {\Omega}_m^{(\dft{h})}}{E_sN}\ma{I}_K \right)^{-1}. \label{eq: rx pulse design}
\end{equation} 
Thus, \small $\IndexM{\ma{W}_{\text{rx}}}{k}{m} = \IndexM{\ma{\Lambda}^{(\text{rx})}_{m}}{k}{k}$\normalsize.
For each frame, the receiver window needs to be updated based on the channel coefficients. First $\{\Omega_m^{(\dft{h})}\}$ are computed along the \ac{ZF}-\ac{CEq}. Assuming the absolute values of $\IndexM{\ma{\Lambda}^{(\text{rx})}_{m}}{k}{k}$ are stored in advance,  only a real-valued reciprocal block and additional complex multiplier are  required to compute the window. 
\subsection{SNR analysis}
Assume a \ac{CEq} matrix \small$\dft{\ma{H}}^H_{m,\text{eq}}$ \normalsize and demodulation matrix \small ${\dft{\ma{B}}_m^H =  \frac{1}{K} \DFT{K}\ma{\Lambda}^{(\text{rx})}_m \DFT{K}^H}$\normalsize. After performing \ac{CEq} and the demodulator convolution, as shown in Fig.~\ref{fig:Advanced_FD}, we get the estimate 
\begin{equation}
\small
\hat{\tilde{\ma{d}}}_m = \underbrace{\tilde{\ma{B}}_m^H\dft{\ma{H}}^H_{m,\text{eq}} \ma{\Lambda}^{(\dft{\ma{h}})}_{m} \dft{\ma{A}}_m}_{\dft{\ma{C}}_m} \dft{\ma{d}}_m + \underbrace{\tilde{\ma{B}}_m^H\dft{\ma{H}}^H_{m,\text{eq}}}_{\dft{\ma{E}}_m}\tilde{\ma{v}}_m.
\end{equation}
Following the remaining demodulator steps, namely, transpose then $M$-I\ac{DFT},  we get the $(k,m)$-th estimated symbol as
\begin{equation}
\hat{d}_{k,m} = A_{k,m} d_{k,m} + \bar{z}_{k,m} + \bar{v}_{k,m},\label{eq: individual symbols}
\end{equation}
where $A_{k,m}$ is the overall gain,  $\bar{z}_{k,m}$ is  the \ac{ISI} due to the final $M$-I\ac{DFT}, and $\bar{v}_{k,m}$ is the sum of the \ac{ICI} resulting from \ac{CEq} and  additive noise. The related power equations are 
\begin{equation}
\small
\begin{split}
A_{k,m} &= \frac{1}{M}\sum_{m=0}^{M-1}\IndexM{\tilde{\ma{C}}_m}{k}{k},~ P_{k,m} = E_s |A_{k,m}|^2,
\end{split}
\end{equation}
\begin{equation}
\small
\begin{split}
I^{\text{ISI}}_{k,m} = \Ex{|\bar{z}_{k,m}|^2} &= \frac{E_s}{M}\sum_{m=0}^{M-1}\left|\IndexM{\tilde{\ma{C}}_m}{k}{k} \right|^2 - P_{k,m}. \label{eq:SINR per symbol}
\end{split}
\end{equation}
\begin{equation}
\small
\begin{split}
\Ex{|\bar{v}_{k,m}|^2}
& = \underbrace{\frac{E_s}{M}\sum_{m=0}^{M-1}\sum_{q = 0, q\neq k }^{K-1} \left|\IndexM{\tilde{\ma{C}}_m}{k}{q}\right|^2}_{I^{\text{ICI}}_{k,m}, ~\mbox{ICI power}} \\
&+  \underbrace{\frac{\dft{\sigma}^2}{M^2}\sum_{m=0}^{M-1}\sum_{q=0 }^{K-1}\left|\IndexM{\dft{\ma{E}}_{m}}{k}{q}\right|^2}_{\sigma^2_{k,m},~\mbox{Noise power}}.
\end{split}
\end{equation}
%The $(k,m)$-th symbol \ac{SINR} is calculated as
\begin{equation}
	\mbox{Thereby, ~~~~ }
\small
\text{SINR}_{k,m}(\ma{h}) = \frac{P_{k,m}}{I^{\text{ISI}}_{k,m} + I^{\text{ICI}}_{k,m}+ \sigma_{k,m}^2}.
\end{equation}
Because $\text{SINR}_{k,m}(\ma{h})$ is independent of $m$, all the subsymbols  in the same subcarrier  have an equal \ac{SINR}. A low complexity computation of the \acp{SINR} values is achieved by exploiting  the circular matrices \small$\{\dft{\ma{A}}_m\}$ and $\{\dft{\ma{B}}_m\}$ \normalsize used to compute \small $\{\ma{C}_m\}$ and $\{\ma{E}_m\}$\normalsize, especially for the practical diagonal \ac{CEq}.
\begin{figure*}[t]
	\centering
	\begin{subfigure}[b]{0.31\textwidth}
		\includegraphics[width=\textwidth]{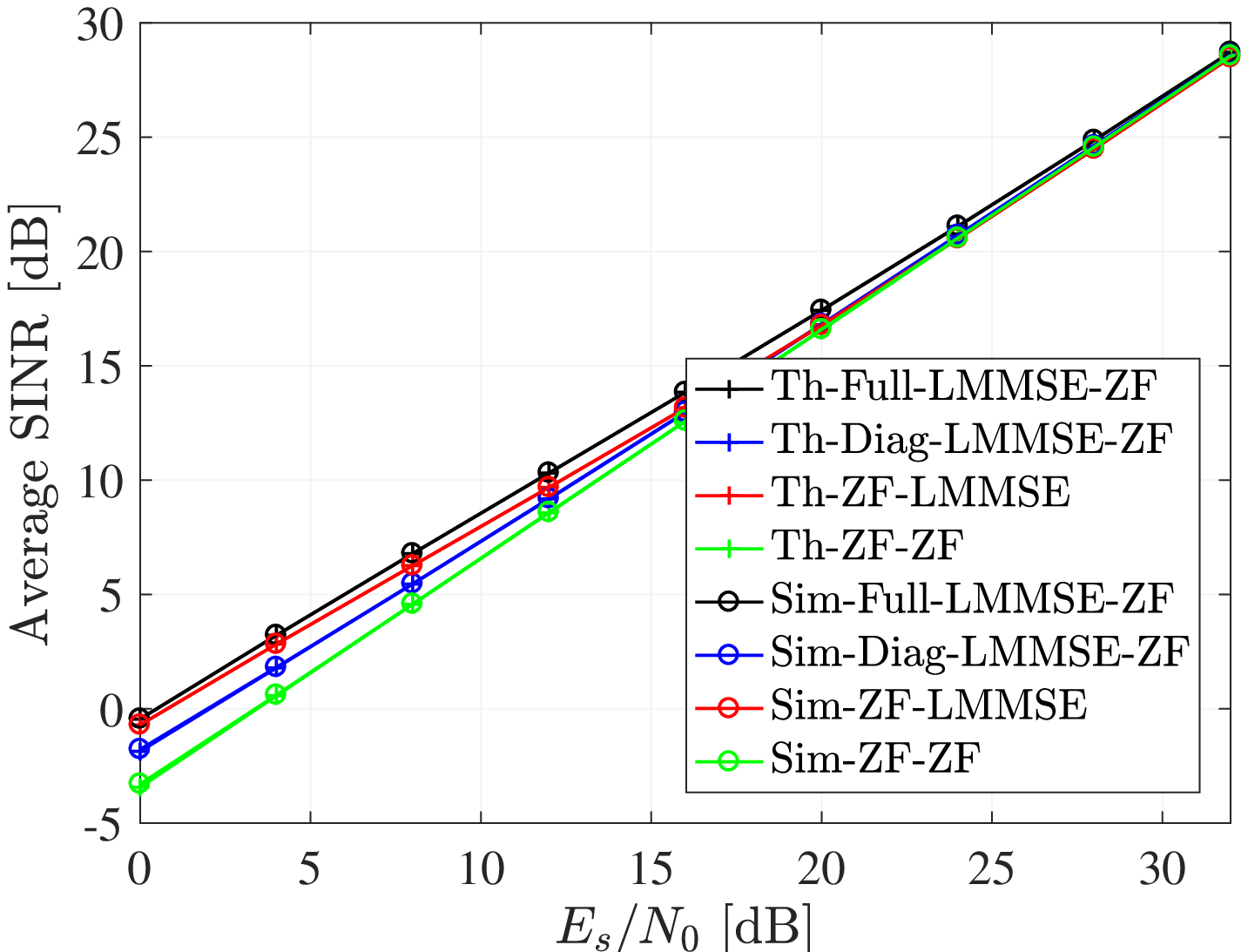}
		\caption{Average symbols \acs{SINR} for $\alpha = 0.8$.}
		\label{fig:conv SINR sim, ana}
	\end{subfigure}
	~ %add desired spacing between images, e. g. ~, \quad, \qquad, \hfill etc. 
	\begin{subfigure}[b]{0.31\textwidth}
		\includegraphics[width=\textwidth]{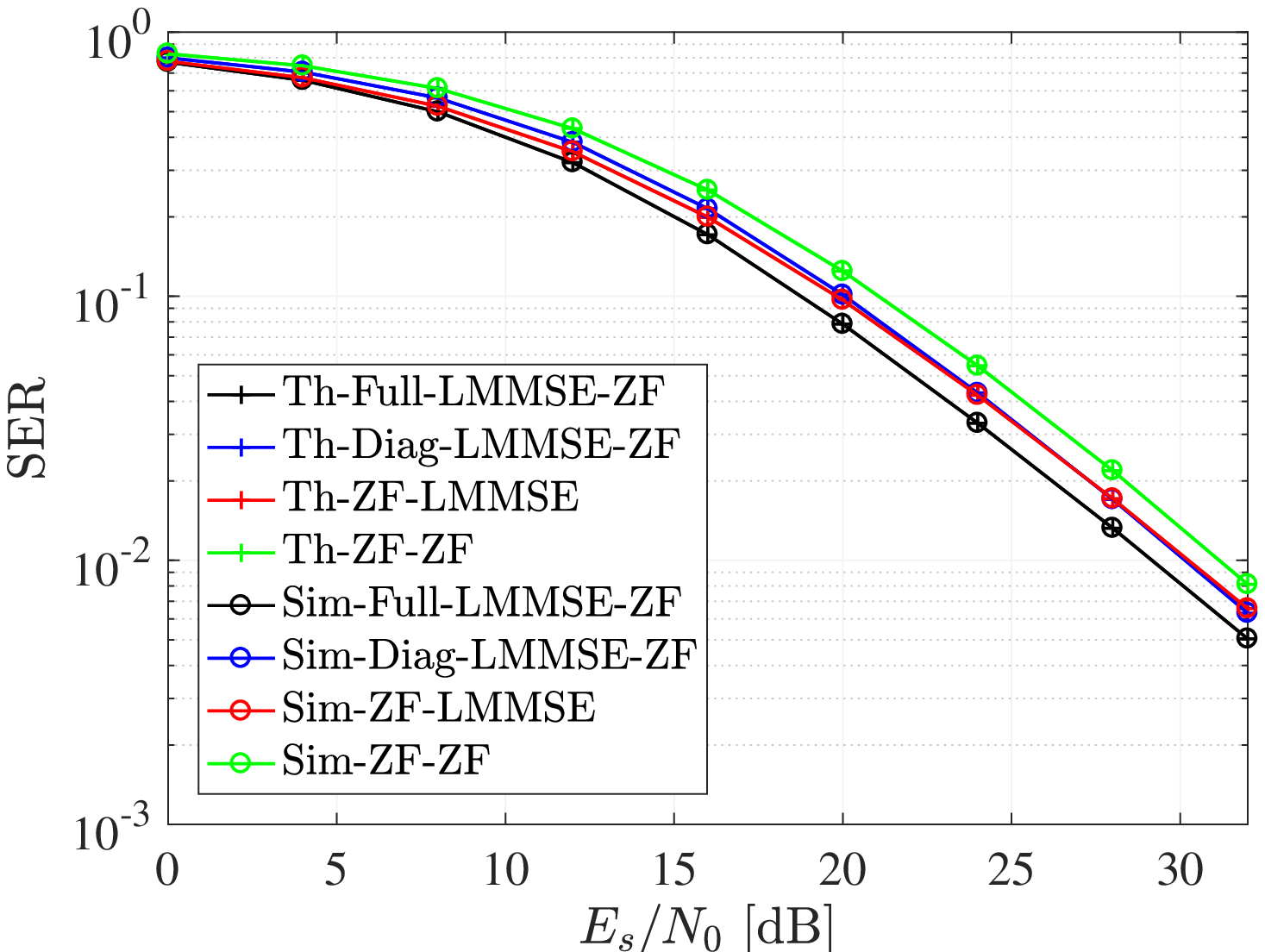}
		\caption{\acs{SER} for $\alpha = 0.8$. }
		\label{fig:conv SER sim, ana}
	\end{subfigure}
	~ %add desired spacing between images, e. g. ~, \quad, \qquad, \hfill etc. 
	\begin{subfigure}[b]{0.31\textwidth}
		\includegraphics[width=\textwidth]{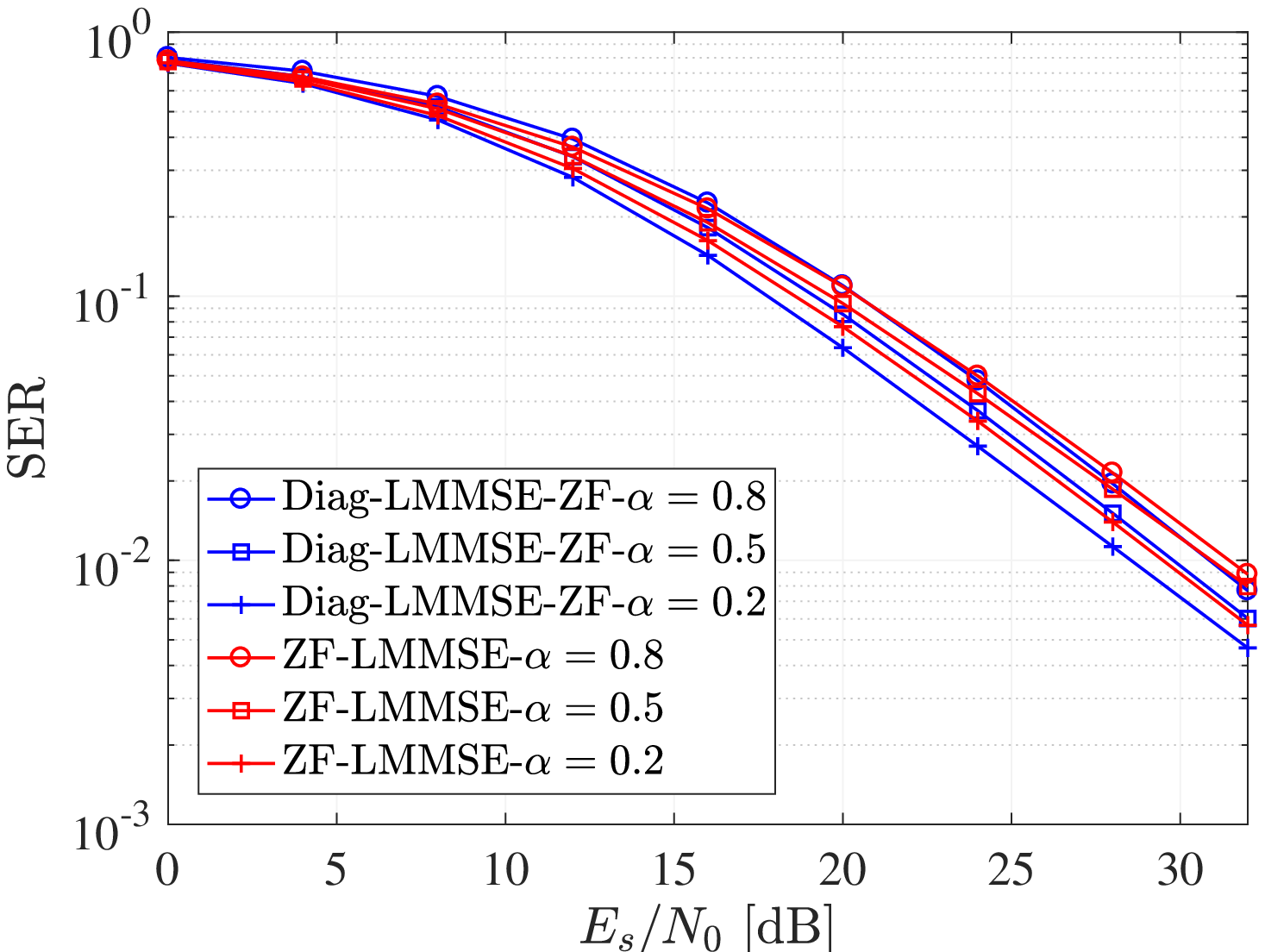}
		\caption{\acs{SER} for different roll-off factors.}
		\label{fig:conv SER roll-off}
	\end{subfigure}
	\caption{Evaluation of non-orthogonal conventional \acs{GFDM} in block fading channel of exponential \acs{PDP}.}\label{fig:non-orhtogonal evaluation}
\end{figure*}
\begin{figure*}
	\centering
	\begin{subfigure}[b]{0.31\textwidth}
		\includegraphics[width=\textwidth]{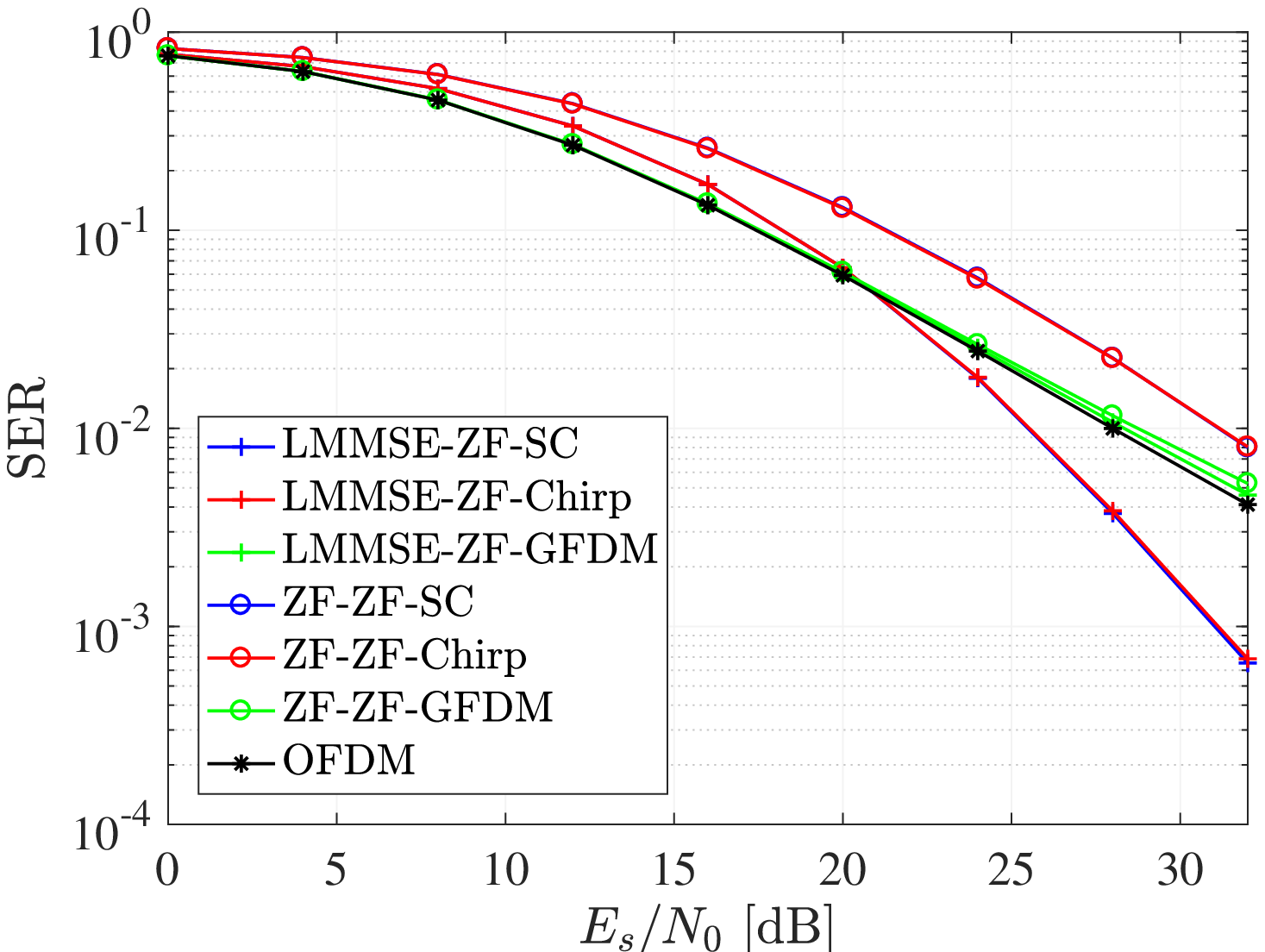}
		\caption{\acs{SER} with exponential \acs{PDP}.}
		\label{fig:SER Exp}
	\end{subfigure}
	~ %add desired spacing between images, e. g. ~, \quad, \qquad, \hfill etc. 
	\begin{subfigure}[b]{0.31\textwidth}
		\includegraphics[width=\textwidth]{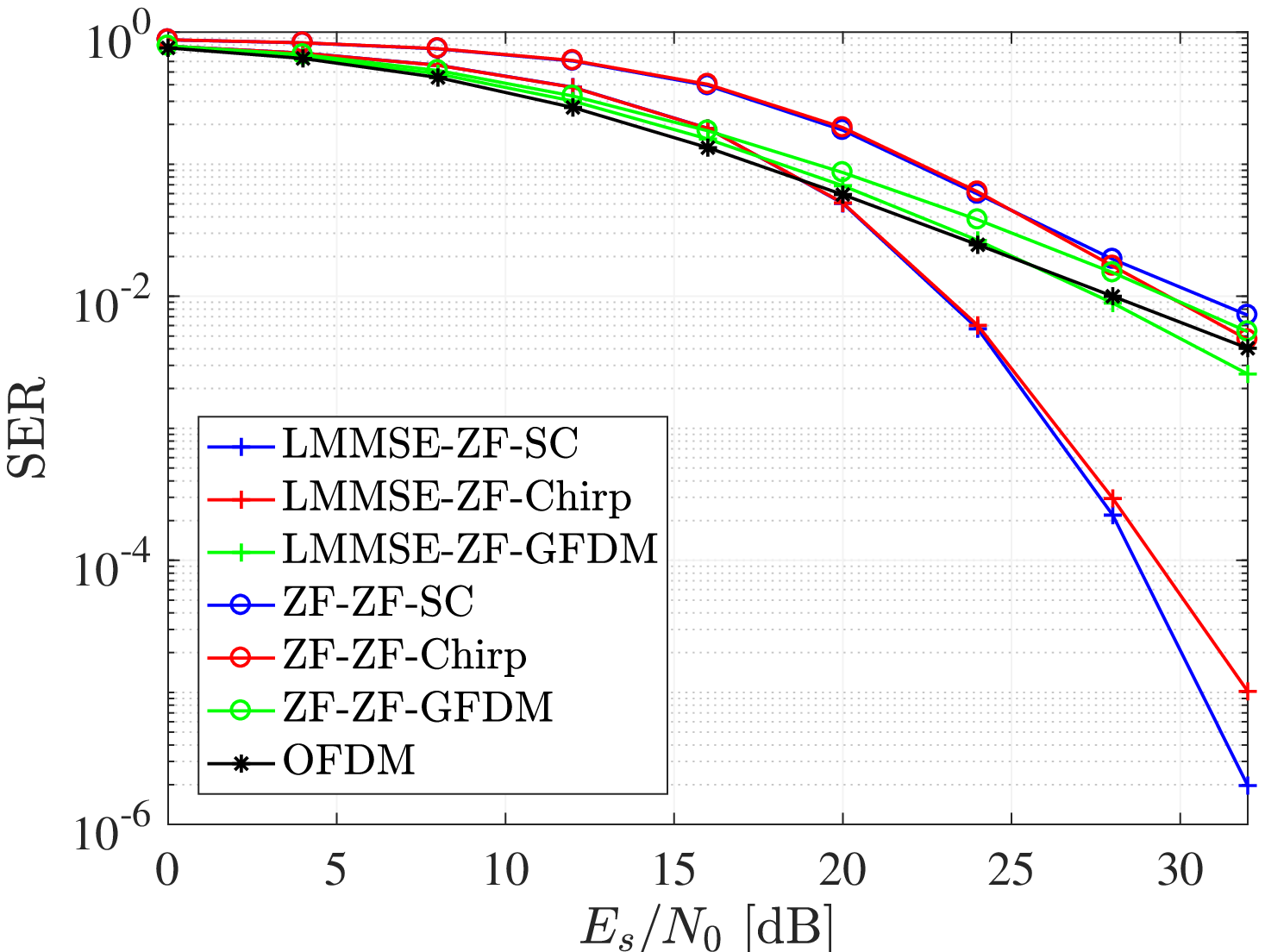}
		\caption{\acs{SER} with uniform \acs{PDP}.}
		\label{fig:SER Uni}
	\end{subfigure}
	~ %add desired spacing between images, e. g. ~, \quad, \qquad, \hfill etc. 
	\begin{subfigure}[b]{0.31\textwidth}
		\includegraphics[width=\textwidth]{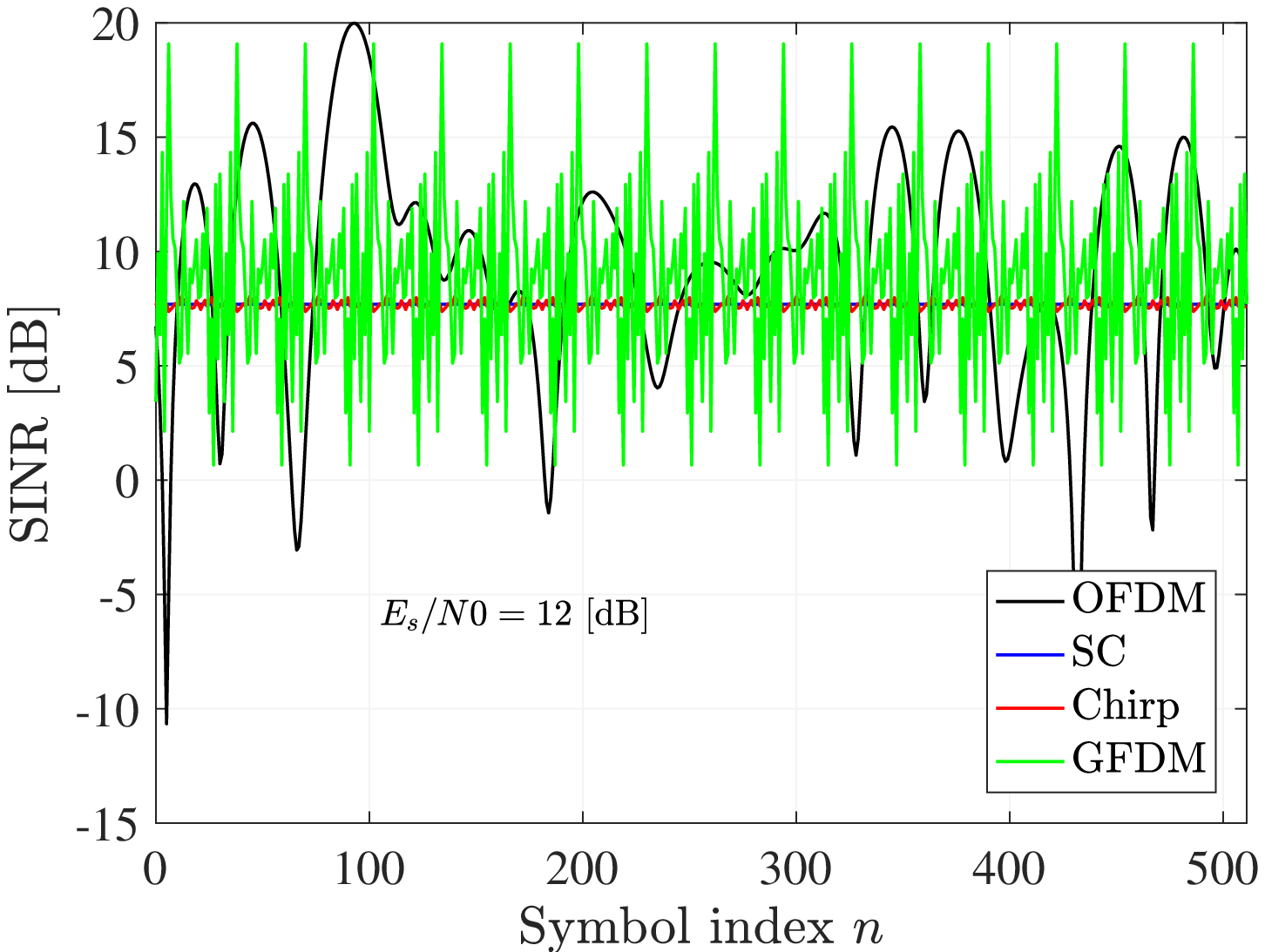}
		\caption{\acs{SINR} per symbol.}
		\label{fig:per symbol power}
	\end{subfigure}
	\caption{Evaluation of orthogonal \acs{GFDM}-based waveforms.}\label{fig:long_f}
\end{figure*}
\section{Evaluation} \label{sec: evaluation}
In this section, we evaluate the performance of the typical \ac{GFDM}-based (\ac{CEq}-Demodulation) receivers in terms of \ac{SER}\footnote{The goal of this simulation is to validate the concept of GFDM-based receiver. For realistic performance evaluation, coded BER should be considered.} for conventional non-orthogonal \ac{GFDM} design and other orthogonal designs based on \ac{GFDM}. 
\subsection{Configurations}
  The conventional non-orthogonal \ac{GFDM} is designed with a  periodic \ac{RC} prototype pulse shape with roll-off factor $0< \alpha <1$, $M\alpha >1$, \cite{nimr2017optimal}. 
The orthogonal case of \ac{GFDM} ($\alpha = 0$) is compared with other orthogonal \ac{GFDM}-based waveforms of the same block length. Namely, \ac{OFDM}, \ac{SC} and Chirp-based. The configuration parameters are listed in Table \ref{tab: design}.
\begin{table}[h!]
	\small
	\caption{Modem  parameters.}\label{tab: design}
\begin{tabular}{l|l|l|l}
&$K$& $M$& Prototype pulse\\
\hline
\rule{0pt}{10pt}GFDM&$32$& $16$ &periodic RC ($\alpha$)\\
\hline
\rule{0pt}{10pt}OFDM&$512$& $1$ & rectangular \ac{TD} pulse\\
\hline
\rule{0pt}{10pt}SC&$1$& $512$ &rectangular \ac{FD} pulse \\
\hline
\rule{0pt}{10pt}Chirp&$32$& $16$ &\small$g[n] = \left\lbrace\begin{array}{cc}
e^{j\pi \frac{n^2}{K}}&,  0\leq n < K\\
0&,  K\leq n < N\\
\end{array}\right\rbrace$\normalsize\\
\hline
\end{tabular}
\end{table}
 We assume the channel impulse response is represented with $L=24$ uncorrelated taps of  \ac{ZMCSC} distribution with exponential or uniform \ac{PDP}. The data symbols are selected from $(M_c = 16)$-QAM constellation  with uniform distribution and mean power $E_s$. The \ac{AWGN} has the variance $N_0$ and the average \ac{SNR} is defined by the ratio $E_s/N_0$. The ($k,m$)-th \ac{SER} for a given channel realization $\ma{h}$ is given by\footnote{$\bar{z}_{k,m} + \bar{v}_{k,m}$ in \eqref{eq: individual symbols} is  Gaussian according to central limit theorem.}\\
\small$\text{SER}_{k,m}(\ma{h}) = 1-\left[1-2\frac{\sqrt{M_c}-1}{\sqrt{M_c}}Q\left(\sqrt{\frac{3\text{SNR}_{k,m}(\ma{h})}{M_c-1}}\right)\right]^2$\normalsize.\\
The  \ac{SER} is computed by averaging over all symbols and channel realizations.
\subsection{Non-orthogonal design}
The possible \ac{CEq} options include Full-\ac{LMMSE} \eqref{eq:CEq lmmase full},   Diag-\ac{LMMSE} \eqref{eq:CEq lmmase diag}, and \ac{ZF}. The demodulation follows the \ac{LMMSE} design \eqref{eq: rx pulse design} or can be \ac{ZF}. The joint \ac{LMMSE} is achieved with the receiver  Full-\ac{LMMSE}-\ac{ZF}. In Fig.~\ref{fig:conv SINR sim, ana} and Fig.~\ref{fig:conv SER sim, ana}, we  verify the closed-form computation of the per-symbol \ac{SINR} \eqref{eq:SINR per symbol} via numerical simulation by illustrating the average \ac{SINR} and \ac{SER} over different channel realizations, respectively. Further, it can be shown that the approximation of joint \ac{LMMSE} outperforms the simple \ac{ZF} receiver. For this particular design, performing \ac{ZF}-\ac{CEq} first approaches the joint \ac{LMMSE} for lower \acp{SNR}. However, when the modulation matrix is well-conditioned, which can be achieved with smaller roll-off factor, the performance of \ac{ZF}-\ac{LMMSE} becomes worse than performing diagonal \ac{LMMSE}-\ac{CEq} first, as shown in Fig.~\ref{fig:conv SER roll-off}. Actually, for $\alpha = 0.2$, the modulation window contains more equal-amplitude columns, as discussed in Section \ref{sec: RX alternative LMMSE demod}. Therefore, the receiver Diag-\ac{LMMSE}-\ac{ZF} is a good approximation of the joint \ac{LMMSE}.  In this context, a hybrid design of the equalizer and the demodulator can be used depending on the $m$-th column of the transmit window. If this  has equal-amplitude values, an exact  diagonal \ac{LMMSE} equalizer is achieved and thus, the corresponding column of the demodulator window is chosen as  \ac{ZF}. In the other case, \ac{ZF} equalization and then \ac{LMMSE} design of the demodulator window is used. On the other hand, it can be seen from Fig.~\ref{fig:conv SER roll-off} that the \ac{SER} decreases with the decrease of $\alpha$, i.e. when the modulation tends to be more orthogonal.
\subsection{Orthogonal design}
 In this section, we compare different orthogonal waveforms with the conventional \ac{GFDM} ($\alpha = 0$).
The optimal joint \ac{LMMSE} is achieved via the diagonal \ac{LMMSE}-\ac{CEq} followed by \ac{ZF}-demodulation. Fig.~\ref{fig:SER Exp}  demonstrates the \ac{SER} for different orthogonal design with exponential \ac{PDP}. In \ac{OFDM},  the data symbols are transmitted over narrow subcarriers, and in \ac{GFDM} over larger subcarriers, whereas in \ac{SC} and chirp-based, they are spread over the whole band. The spreading allows higher frequency diversity, which is observed by the decreased \ac{SER} at higher \acp{SNR}. This gain can be significantly observed with uniform \ac{PDP}, as depicted in Fig.~\ref{fig:SER Uni}. The gap between \ac{ZF} and \ac{LMMSE} is larger in the spreading case. Furthermore, a slightly better performance of \ac{GFDM} is observed compared to \ac{OFDM}. This is because the symbols are spread on wider subcarriers than that of \ac{OFDM}. As a result, the variation in the symbol's \acs{SINR} is smaller. The  equal \ac{SINR} per symbols is attained with \ac{SC} for a given channel realization, as illustrated in Fig.~\ref{fig:per symbol power}. The Chirp-based has a slight variation, whereas \acs{OFDM} suffers from a significant variation in the \acp{SINR}.
\acresetall
\section{Conclusion}\label{sec:conclusions}
In this work, we consider \acs{GFDM}  as a multicarrier framework  to process different waveforms.  For practical implementation, the \acs{GFDM}-based receiver is decoupled into \ac{CEq} and demodulation. The \acs{LMMSE}-\ac{CEq} under diagonal constraint followed by \acs{ZF}-demodulation achieves better performance than the simple \acs{ZF} receiver. Moreover, it approaches the performance of optimal joint \acs{LMMSE} for  non-orthogonal \acs{GFDM} waveform and it becomes exact for orthogonal designs. The alternative \acs{ZF}-\ac{CEq} followed by \acs{LMMSE}-\acs{GFDM} demodulation is more appropriate when the self-interference due to non-orthogonality is higher.  Thereby, a hybrid design of the equalizer and the demodulator can benefit from the structure of the modulator window.  The complexity of the receiver is actually influenced by  the non-orthogonality of the modulation. However,  the performance in terms of \acs{SER} tends to improve when the self-interference is reduced. Thus, the orthogonal modulation achieves better performance with low complexity implementation. Considering orthogonal design, we show that spreading the data symbols over wider subcarriers enables frequency diversity. Accordingly, in block fading frequency selective channels, \ac{SC} and Chirp-based \acs{GFDM} with spreading over the whole bandwidth can achieve very low \acs{BER} employing without channel coding. Additionally, the conventional \acs{GFDM} with wider subcarrier spacing can outperform \acs{OFDM} in certain channels.
\bibliographystyle{IEEEtran}
\bibliography{references}
%%\newpage
\appendix
\subsection{Proof of Lemma 1}\label{sec: MMSE appindx}
For a general linear model \small ${\ma{y} = \ma{G}\ma{d} + \ma{v}}$\normalsize, the \acs{LMMSE} \cite{kay1993fundamentals} receiver matrix can be expressed in two forms as 
\begin{equation}
\small
\begin{split}
\ma{W}^H &= \ma{R}_d\ma{G}^H\left(\ma{G}\ma{R}_d\ma{G}^H+\ma{R}_v\right)^{-1}\\
&= \left(\ma{G}^H\ma{R}_v^{-1}\ma{G}+\ma{R}_d^{-1}\right)^{-1}\ma{G}^H\ma{R}_v^{-1}.
\end{split} \label{eq: MMSE}
\end{equation}
Let \small $\ma{G} = \ma{H}\ma{A}$, $\ma{R}_d = E_s\ma{I}_N$\normalsize, and \small $\ma{R}_v  = \sigma^2\ma{I}_N$\normalsize, the \acs{LMMSE} using the first line of \eqref{eq: MMSE}  is given by
\begin{equation*}
\small
\begin{split}
\ma{W}^H &=  \underbrace{{E_s}\ma{A}^H\left({E_s}\ma{A}\ma{A}^H+{\sigma^2}\ma{H}^{-1}{\ma{H}^{H}}^{-1}\right)^{-1}}_{\ma{B}^H_{\text{LMMSE}}} \ma{H}^{-1}.
\end{split}
\end{equation*}
Noting that, \small $\sigma^2\ma{H}^{-1}{\ma{H}^{H}}^{-1} = \Ex{\bar{\ma{v}}\bar{\ma{v}}^H}$, $\bar{\ma{v}} = \ma{H}^{-1}\ma{v}$\normalsize. Then, $\ma{B}^H_{\text{MMSE}}$ is the \acs{LMMSE} demodulation matrix with respect to \small $\ma{y}_{\text{zf}} = \ma{A}\ma{d} + \bar{\ma{v}}$\normalsize. As a result, \small$\hat{\ma{d}} = \ma{W}^{H}\ma{y} = \ma{B}^H_{\text{LMMSE}}\left[\ma{H}^{-1}\ma{y}\right]$\normalsize.
On the other hand, by using the second line of \eqref{eq: MMSE}
\begin{equation*}
\small
\begin{split}
\ma{W}^H &=  \ma{A}^{-1}\underbrace{\left(\frac{1}{\sigma^2}\ma{H}^H\ma{H}+[{E_s}\ma{A}\ma{A}^H]^{-1}\right)^{-1}\ma{H}^H\frac{1}{\sigma^2}}_{\ma{H}_{\text{LMMSE}}}.
\end{split}
\end{equation*}
Here, \small$E_s \ma{A}\ma{A}^H = \Ex{\bar{\ma{d}}\bar{\ma{d}}^H}$,~ $\bar{\ma{d}} = \ma{A}\ma{d}$\normalsize. Thus, $\ma{H}_{\text{LMMSE}}$ is the \acs{LMMSE} channel equalization with respect to $\ma{y} = \ma{H}\bar{\ma{d}}+\ma{v}$. Accordingly, \small$\hat{\ma{d}} = \ma{W}^{H}\ma{y} = \ma{A}^{-1}\left[\ma{H}_{\text{LMMSE}}~ \ma{y}\right]$\normalsize.

\end{document}